# Binary Prime Tableau Sequences

Prashanth Busireddygari and Subhash Kak

Oklahoma State University, Stillwater, OK, USA

**Abstract.** This paper proposes a new class of random sequences called binary primes tableau (PT) sequences that have potential applications in cryptography and communications. The PT sequence of rank $p$ is obtained from numbers arranged in a tableau with $p$ columns where primes are marked off until each column has at least one prime and where the column entries are added modulo 2. We also examine the dual to the PT sequences obtained by adding the rows of the tableau. It is shown that PT sequences have excellent autocorrelation properties.

## 1  Introduction

Random sequences have many applications in programming [Knuth. 1997], cryptography [Luby. 1996] and communications [Golomb and Gong 2005] and they are useful in simulation experiments. The physical basis of randomness is an important problem [Landauer. 1996] and it takes us back to the very nature of information in quantum theory [Kak. 1999, Kak. 2007]. The question of randomness is also fundamental in cognitive science and decision theory [Aerts. 2009, Kak. 1996].

From a mathematical perspective, pseudo-randomness may be related to the complexity of certain number-theoretic properties [Cassaigne et al. 1999, Mauduit. 2002]. Specifically, one may seek to define the unexpectedness of some properties of composition [Chen. 1978, Kak. 2014, Nicolas and Robin 1997] that yield good randomness measure. In a recent paper, we showed that binary primes sequence can be used for computational hardening of pseudorandom sequences [Reddy and Kak 2016]. Here we go beyond that idea and generate pseudorandom sequences by folding the binary primes sequences using p columns.

This is done by starting with numbers ordered in a tableau with a prime number of elements in each row. Next all composite numbers are marked as 0 and the primes themselves are marked as 1. We stop the process once each column has a 1. The columns are added mod 2 to yield a 0 or 1 in each position. If the rows are added we get the dual sequence.

We show in this paper that such tableau sequences have outstanding autocorrelation properties and, therefore, they can be used in many cryptography and communications applications.

## 2 The Numbers Tableau

**Definition 2.1** *The Numbers Tableau ($NT_p$) of rank p is a table with p columns listed as 0 to p-1, where the entries in the tableau in the kth row are the natural numbers from (k-1)p through (kp-1).*

$$\begin{vmatrix} 0 & \ldots & \ldots & \ldots & p-1 \\ p & \ldots & \ldots & \ldots & 2p-1 \\ \vdots & \vdots & \vdots & \ddots & \vdots \\ (k-1)p & \ldots & \ldots & \ldots & kp-1 \end{vmatrix}$$

$NT_p$ is a way of representing integers in $p$ columns.

**Example 2.1** Let $p = 3$ and $k = n$ (number of rows). The $NT_p$ for $p$ is therefore

$$\begin{vmatrix} 0 & 1 & 2 \\ 3 & 4 & 5 \\ 6 & 7 & 8 \\ \vdots & \vdots & \vdots \\ (n-1)p & \ldots & np-1 \end{vmatrix}$$

**Definition 2.2** *The Binary Primes Tableau ($BPT_p$) is a table with p columns listed as 0 to p-1, where the entry is 0 if the number is composite and 1 if it is prime.*

**Example 2.2** Let $p = 5$ and $k = n$ (number of rows). The $BPT_p$ for $p$ is given by

$$\begin{vmatrix} 0 & 0 & 1 & 1 & 0 \\ 1 & 0 & 1 & 0 & 0 \\ 0 & 1 & 0 & 1 & 0 \\ \vdots & \vdots & \vdots & \vdots & \vdots \\ \ldots & \ldots & \ldots & \ldots & \ldots \end{vmatrix}$$

# 3 The Binary Primes Tableau Sequence

**Definition 3.1** *Complete Binary Primes Tableau ($CBPT_P$): $CBPT_P$ is a finite termination of the $BPT_P$ when there is a 1 in each of the columns of the $BPT_P$.*

There are two options for this termination which lead to what we call sequences of the first kind and the second kind. First, the remaining entries in the last row are put equal to zero. Such a termination will be used in the remainder of the paper. The second option is to let the sequence continue until the end of the row.

**Kind 1** Let $p = 7$. The $CBPT_P$ for $p$ is constructed as

$$\begin{vmatrix} 0 & 0 & 1 & 1 & 0 & 1 & 0 \\ 1 & 0 & 0 & 0 & 1 & 0 & 1 \\ 0 & 0 & 0 & 1 & 0 & 1 & 0 \\ 0 & 0 & 1 & 0 & 0 & 0 & 0 \\ 0 & \boxed{1} & & & & & \end{vmatrix}$$

The one marked by a square in this matrix fulfills the condition that each column have 1.

**Kind 2 1** Let $p = 7$. The $CBPT_P$ for $p$ is constructed as

$$\begin{vmatrix} 0 & 0 & 1 & 1 & 0 & 1 & 0 \\ 1 & 0 & 0 & 0 & 1 & 0 & 1 \\ 0 & 0 & 0 & 1 & 0 & 1 & 0 \\ 0 & 0 & 1 & 0 & 0 & 0 & 0 \\ 0 & 1 & 0 & \boxed{1} & 0 & 0 & 0 \end{vmatrix}$$

**Definition 3.2** *Binary Primes Tableau Sequence (BPT): The sequence obtained by adding $CBPT_P$ entries column-wise mod 2, will be called a Binary Primes Tableau sequence.*

**Theorem 3.1** *As p becomes large, the BPT sequence is random from an information complexity point of view.*

**Proof 3.1** If $i$ is a non-zero entry in each column then by the prime number theorem the probability of finding the location of the non-zero entries ordered by rank $p$ for numbers up to $N_{max}$ in $q$ rows by the use of a sieve is $O(p^{pq})$. As $N_{max} \to \infty$, the total number of operations rises uncountably and so the sequence is random from an information-theoretic complexity.

$$\prod_{i=1}^{\infty}\binom{p}{i} \tag{1}$$

where $\sum_{i=1}^{\infty} = \dfrac{pq}{\log pq}$

The *BPT* may be related to partitions via the Stirling number of the second kind. A Stirling number measures the ways to partition a set of $n$ objects into $k$ non-empty subsets and it is denoted by $S(n, k)$ or $\binom{n}{k}$. The $k$ subsets here may be compared to the number of columns and $n$ the number of primes that are being considered until each column is filled.

We know that

$$\sum_{k=0}^{n}\binom{n}{k} x(x-1)(x-2)\ldots(x-n+1) = x^n \tag{2}$$

The value of $\binom{n}{k}$ increases in a geometric fashion as $k$ increases.

$$\binom{n}{k} = \sum_{j=1}^{k}(-1)^{k-j}\frac{j^{n-1}}{(j-1)!(k-j)!} = \frac{1}{k!}\sum_{j=1}^{k}(-1)^{k-j}\binom{k}{j}j^n \tag{3}$$

However the two cases are not identical because we do not have an exact count of the primes needed to complete the tableau. But the fact that the Stirling numbers increases rapidly highlights the information complexity aspect of the *BPT* sequence.

**Definition 3.3** *The Dual Tableau sequence (DT): The dual to the BPT is the sequence that is obtained by adding $CBPT_P$ entries row-wise mod 2.*

For the same reasoning as in **3.1**, one concludes that the dual tableau sequence is also random as $n_{max}$ becomes large.

**Theorem 3.2** *As p becomes large, the DT sequence is random from an information complexity point of view.*

**Example 3.1** Let $p = 13$. The *BPT* and *DT* for $p$ is constructed as

```
0 0 1 1 0 1 0 1 0 0 0 1 0    1
1 0 0 0 1 0 1 0 0 0 1 0 0    0
0 0 0 1 0 1 0 0 0 0 0 1 0    1
0 0 1 0 1 0 0 0 1 0 0 0 0    1
0 1 0 0 0 0 0 1 0 1 0 0 0    1   Dual Tableau Sequence (DT)
0 0 1 0 0 0 1 0 1 0 0 0 0    1
0 1 0 0 0 1 1 0 0 0 0 1 0    1
0 0 0 0 0 0 1 0 0 0 1 0 1    1

1 0 1 0 0 1 1 0 0 1 0 1 1
```

Binary Primes Tableau Sequence (BPT)

In this example the dual is not balanced and therefore, it does not behave random. But as $p$ increased the sequence becomes random for exactly the same reason as outlined in the proof of **3.1**

**Definition 3.4** $N_{max}$: *The prime number for which the last entry becomes non-zero will be called $N_{max}$ and the ratio $\frac{N_{max}}{p}$ will be called as chunk size.*

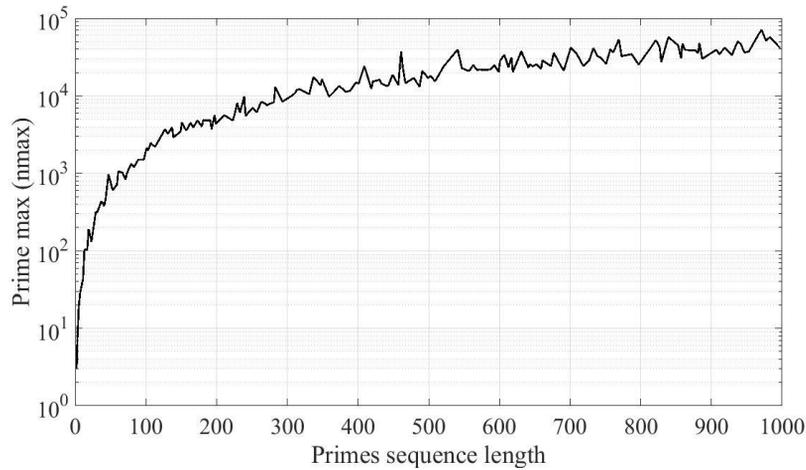

**Fig. 1.** $N_{max}$ variation

Let the weight of the sequence be $k$. Then the ratio $\frac{k}{n}$ gives a measure of balance. If $\frac{k}{n} = \frac{1}{2}$, then the sequence is perfectly balanced.

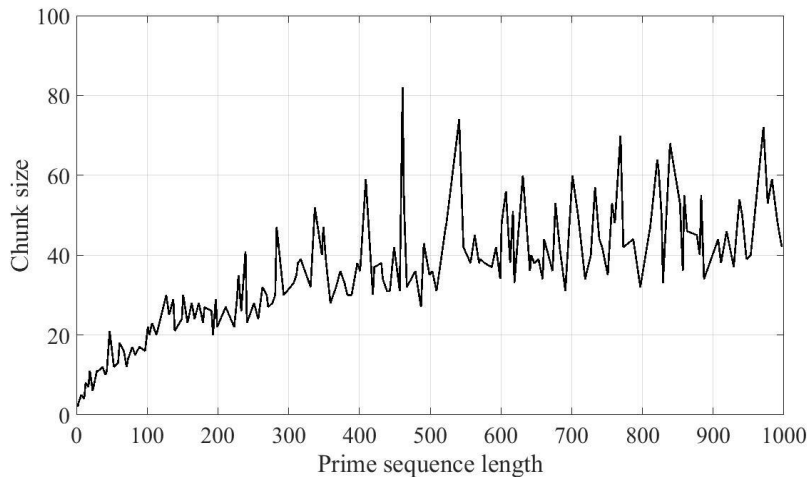

**Fig. 2.** Chunk size, $(\frac{N_{max}}{p})$ variation

**Table 1.** The measure of balance, $\frac{k}{n}$ in *BPT* and *DT* sequence

| Prime | BPT | DT |
|-------|-------|-------|
| 13 | 0.461 | 0.125 |
| 199 | 0.532 | 0.681 |
| 461 | 0.527 | 0.672 |
| 971 | 0.522 | 0.465 |
| 997 | 0.534 | 0.523 |

As we see the Table 1, as *p* increases the ratio of 1s and 0s in both *BPT* and *DT* become approximately equal.

## 3.1 Properties of the BPT sequence and the Dual sequence

Many of the commonly used pseudorandom number generators (PRNGs) generate sequences of shorter periods for a given seed. They lack balanced distribution of numbers which later effects the correlation properties and randomness of the sequence.

The *BPT* sequence generated has balanced distribution of 0s and 1s. The ratio of 1s and 0s are naturally balanced and for a given BPT sequence length the occurrence of 1s is approximately equal to 53% for the range of primes that we have tested.

The range of the last non-zero prime number, $N_{max}$ in the $NT_P$ to generate the *BPT* sequence is unpredictable and remains as a challenge for the attacker to find.

The maximum periodic length of the *BPT* sequence is fixed depending on the tableau size and has only one single cycle.

## 4 Correlation Properties

The mean square periodic autocorrelation (MSPAC) and mean square periodic crosscorrelation (MSPCC) are useful to study the randomness properties of PN sequences. PN sequences with low autocorrelation and crosscorrelation measurements offers a great difficulty for the attackers to guess [Ziani and Medouri 2015].

We calculate the periodic correlation function for a sequence by shifting the non-delayed version of sequence with the delayed version cyclically and is defined as:

$$r_{i,j}(\tau) = \frac{1}{N} \sum_{i=1}^{N-1} c_i(n) * c_j(n+\tau) \qquad (4)$$

where $c_j(n+\tau)$ represents the delayed version of maximum length sequence $c_i$ by $\tau$ units, and $c_i$ represents the non-delayed version of $c_i$.

The MSPAC measurement is given by:

$$\frac{1}{M} \sum_{i=1}^{M} \sum_{\tau=1-N, \tau \neq 0}^{N-1} |r_{i,j}(\tau)|^2 \qquad (5)$$

There exists multiple ways to compute the crosscorrelation function of a sequence. One way is to compute it by taking the L.C.M. of the two sequences of different length. In such computation, we let the shorter sequence run through multiple periods until it match the length of the larger sequence. We rely upon that computational procedure to calculate the crosscorrelation of PN sequences in this paper.

The MSPCC measurement is defined as:

$$\frac{1}{M(M-1)}\sum_{i=1}^{M}\sum_{j=1,j\neq i}^{M}\sum_{r=1-N}^{N-1}|r_{i,j}(\tau)|^2 \qquad (6)$$

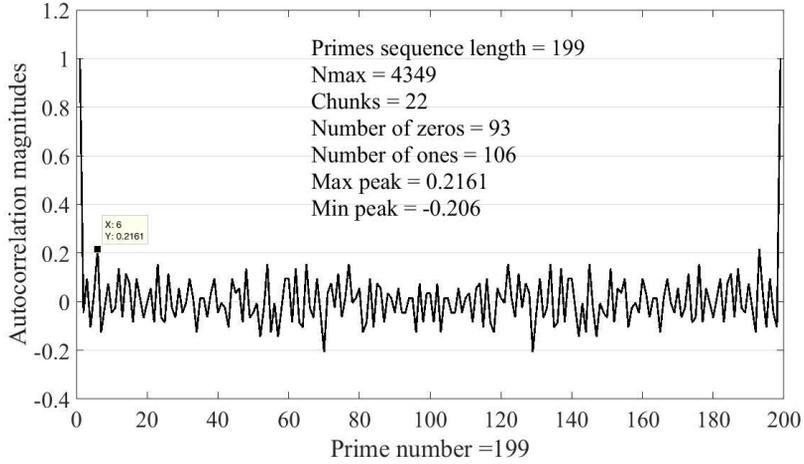

**Fig. 3.** Autocorrelation function for a sequence of length 199

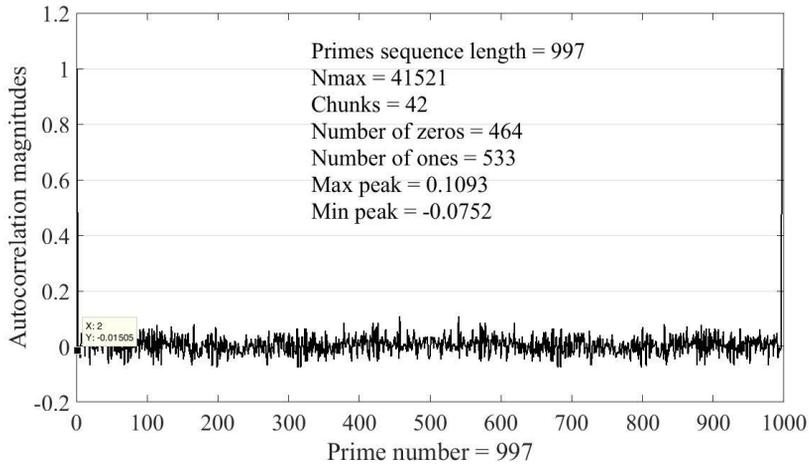

**Fig. 4.** Autocorrelation function for a sequence of length 997

Table 2 presents the comparison of autocorrelation and crosscorrelation peaks measurements of *BPT* sequence described in the paper and other classes that are established in the literature. For all the sequences, the codes length has been taken as 199 bits.

**Table 2.** Correlation measures for various pseudorandom sequences

| Sequence | Autocorrelation | Crosscorrelation |
|---|---|---|
| BPT | 0.1161 | 0.2052 |
| Gold | 0.646 | 0.828 |
| Small Kasami | 0.547 | 0.766 |
| Large Kasami | 0.832 | 0.601 |

From the Table 2, we notice *BPT* has the best MSPAC and MSPCC measurement compared to other sequences.

## 5  Conclusions

This paper proposed a new class of random sequences called binary primes tableau (*BPT*) sequences that have potential applications in cryptography and communications. The *BPT* sequence of rank *p* is obtained from numbers arranged in a tableau with *p* columns where primes are marked off until each column has at least one prime and where the column entries are added modulo 2. We investigated the dual to the PT sequences obtained by adding the rows of the tableau. It was shown that *PT* sequences have excellent autocorrelation properties.